\documentclass[aps,prl,twocolumn,a4paper,reprint,longbibliography,groupedaddress,floatfix,superscriptaddress]{revtex4-2}

\usepackage[dvipsnames]{xcolor}
\definecolor{linkcolor}{rgb}{0.3,0.3,1.0} %hyperlink
\usepackage[pdftex,colorlinks=true, linkcolor= linkcolor, citecolor= linkcolor, urlcolor= linkcolor, hyperindex=true,hyperfigures=true]{hyperref} %hyperlink

\usepackage{amssymb,amsmath,amsfonts}
\usepackage{graphicx}
\usepackage{float}
\usepackage{bm}
\usepackage{dcolumn}%

\begin{document}

\title{Inter-event time statistics of earthquakes as a gauge of volcano activity}

   \author{Sumanta Kundu}
   \email[ ]{skundu@sissa.it}
   \affiliation{
   \begin {tabular}{c}
    Department of Earth and Space Science, Osaka University, 560-0043 Osaka, Japan
   \end {tabular}}
   \affiliation{Scuola Internazionale Superiore di Studi Avanzati (SISSA), Via Bonomea 265, 34136 Trieste, Italy}

   \author{Anca Opris}
   \affiliation{
   \begin {tabular}{c}
    Department of Earth and Space Science, Osaka University, 560-0043 Osaka, Japan
   \end {tabular}}
   
    \author{Yosuke Aoki}
   \affiliation{
   \begin {tabular}{c}
    Earthquake Research Institute, The University of Tokyo, 113-0032 Tokyo, Japan
    \end {tabular}}

   \author{Takahiro Hatano}
   \email[ ]{hatano@ess.sci.osaka-u.ac.jp}
   \affiliation{
   \begin {tabular}{c}
    Department of Earth and Space Science, Osaka University, 560-0043 Osaka, Japan
   \end {tabular}}

%\date{\today} 
\begin{abstract}
The probability distribution of inter-event time (IET) between two consecutive earthquakes is a measure for the uncertainty in the occurrence time of earthquakes in a region of interest. It is well known that the IET distribution for regular earthquakes is commonly characterized by a power law with the exponent of 0.3. However, less is known about other classes of earthquakes, such as volcanic earthquakes, which do not manifest mainshock-aftershocks sequences. Since volcanic earthquakes are caused by the movement of magmas, their IET distribution may be closely related to the volcanic activities and therefore of particular interest. Nevertheless, the general form of IET distribution for volcanic earthquakes and its dependence on volcanic activity are still unknown. Here we show that the power-law exponent characterizing the IET distribution exhibits a few common values depending on the stage of volcanic activity. Volcanoes with steady seismicity exhibit the lowest exponent ranging from 0.6 to 0.7. During the burst period, when the earthquake rate is highest, the exponent reaches its peak at approximately 1.3. In the preburst phase, the exponent takes on the intermediate value of 1.0. These values are common to several different volcanoes. Since the preburst phase is characterized by the distinct exponent value, it may serve as an indicator of imminent volcanic activity that is accompanied by a surge in seismic events.
\end{abstract}

\maketitle

%%%%%%%%%%%%%%%%%%%%%%%%%%%%%%%%%%%%%%%%%%%%%%%%%%%%%%%%%%%%%%%%%%%%%%%%%%%%%%%%%%%
\noindent\textit{Introduction - }Occurrence of earthquakes appear to be unpredictable, but their statistics exhibits some reproducible and universal features. Among them, the Gutenberg-Richter (GR) law and the Omori law have marked a milestone in statistical seismology. A statistical model for earthquakes, known as the epidemic-type aftershock sequence model, is essentially based on these two empirical laws and has been widely recognized as a standard model in this field of study~\cite{Ogata2017}.
 
 On the other hand, there have been some other classes of earthquakes that exhibit peculiar statistical properties. It is well known that numerous earthquakes can occur in a relatively small region without apparent mainshocks and continue for a certain duration, typically a few months. This phenomenon is referred to as a swarm. The GR law generally holds for swarms, but the Omori law does not hold since swarms do not have apparent mainshocks and aftershocks~\cite{Horalek2021}.
 
 The peculiar statistics of swarms may be due to their physical origin. It is believed that swarms are driven by the stress change caused by geofluids~\cite{Horalek2008,Shelly2013,Brauer2009}, since the ratio of longitudinal to transverse wave velocities, denoted by $V_P/V_S$, is generally large in the proximity of swarm hypocenters. This is quite different from ordinary earthquakes, which are driven by slow and secular tectonic stressing.
     
 Swarms are often observed in volcanic regions. In particular, earthquakes occurring at shallow depth in the close proximity of active volcano are defined as volcanic earthquakes. They also do not exhibit mainshock-aftershocks sequence~\cite{Traversa2010} and therefore may be classified as a swarm. The main driving force of volcanic earthquakes is the intrusion of ascending magma, typically a form of dike, into a volcano. Namely, the pressurized magmatic fluids beneath the volcanic edifices cause the fracturing of brittle rocks~\cite{Chouet2003,McNutt_VT2005,Roman2006,nishimura_VT2011}. 
 
 The GR law holds for volcanic earthquakes as well. However, the b-value can be larger than typical values for ordinary earthquakes~\cite{Roberts2015}. This may be due to the difference in physical mechanism of fracture as explained above.
 
 The transport processes of geofluids are generally much faster than tectonic loading timescale. In addition, fluids can affect the stress states of faults mostly through normal stress, since fluids cannot support shear stress. Therefore, the stress evolving process should be quite different between swarms and ordinary earthquakes.
  
%  For instance, Matsushiro and Noto cases [{\bf References?} Yoshida et al. 2023]. In such cases, the depth of hypocenters can be as deep as $15$ km, not limited to a shallow part.  To support this hypothesis, earthquakes induced by hydrofracking also occur as swarms [{\bf References?}]
% Yoshida, K. et al. (2023) Upward earthquake swarm migration in the northeastern Noto Peninsula, Japan, initiated from a deep ring-shaped cluster: Possibility of fluid leakage from a hidden magma system. Journal of Geophysical Research: Solid Earth, 128, e2022JB026047.
   
   % Why IET
   Since the Omori law is based on a concept of aftershocks, it cannot be applied to these peculiar classes of earthquakes, which do not exhibit mainshock-aftershocks sequence. One should thus study more general quantity that characterizes the statistical properties of seismic time series.
   
   Here we study the statistics of inter-event time to unveil the statistical properties of volcanic earthquakes. The inter-event time (IET) is the duration between two consecutive earthquakes in a given area. Therefore it is clearly defined without any definitions of mainshocks and aftershocks. The probability distribution of IET characterizes the temporal properties of point process, and encompasses the full temporal information if the point process is a renewal process.
   % For instance, the IET distribution is exponential for Poissonian process.
   % Other example?   

   The IET distribution has been investigated extensively for ordinary earthquakes~\cite{Corral2004,Corral2006,Davidsen2013}. Irrespective of the area/region or the time window (stationarity of seismicity), the IET distribution is well described by the Gamma distribution:
\begin{equation}
\label{eq:GammaDistribution}
    p(\tau) = a\tau^{-\alpha} e^{-\tau/b}.
\end{equation}
   This essentially means that the IET distribution follows a power-law decay $p(\tau) 
   \sim \tau^{-\alpha}$ at intermediate time scale. The exponent $\alpha$ appears to have a robust value $\approx 0.3$ for ordinary earthquakes.
   The IET distribution for volcanic earthquakes has been investigated by Lehr et al. for a volcano (Villarrica) in Chile, but they cannot draw decisive conclusion on the functional form of IET distribution due to the limited number of volcanic earthquakes~\cite{Lehr2020}.
   
   In this paper, analyzing data from seven volcanoes (see catalog details in Table \ref{tab:catalog}), we show that a relation similar to Eq. (\ref{eq:GammaDistribution}) holds for volcanic earthquakes and discuss how the exponent value $\alpha$ depends on the region, the time window, and the corresponding volcanic seismicity.

\textit{Results - }In order to test whether and to what extent the IET distribution depends on the long-term volcanic activity and how they are different from the tectonic seismicity, we consider two broad classes of time series: the stationary and the non-stationary volcanic seismicity.

\textit{Stationary time series - }
%
%===============================================
    \begin{figure}[t]
    \centering
    \includegraphics[width=0.8\linewidth]{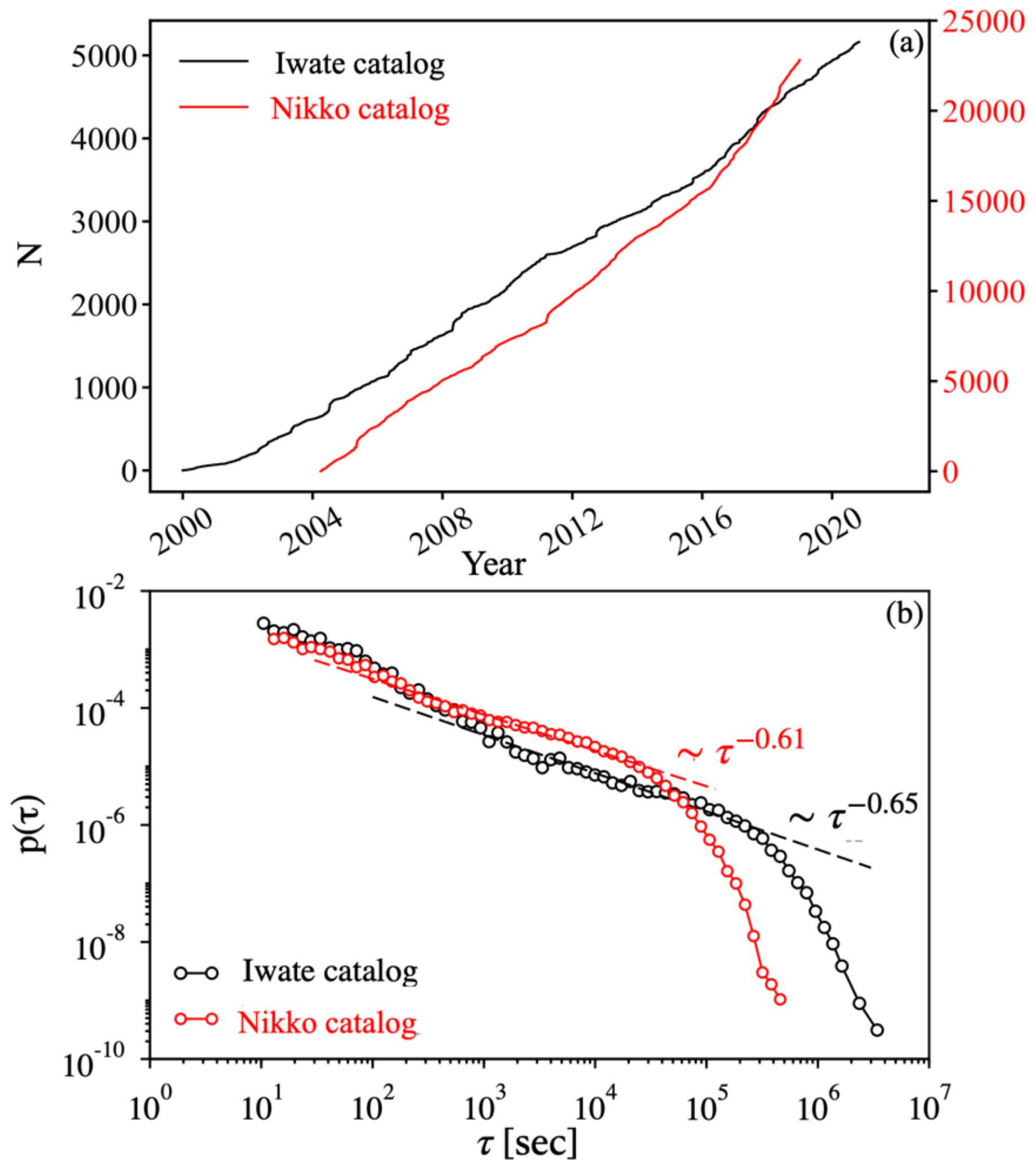}
    \caption{{\bf The characteristic features of the stationary volcanic time series. (a)} The cumulative number of events as a function of time for the volcanic seismicity recorded at Iwate and Nikko within the observation window with $M \geqslant M_{\text{min}}$. The linear trend of the curves indicating the stationary seismicity. {\bf(b)} Log-log plot of the probability density distributions $p(\tau)$ of the inter-event times $\tau$. The dashed lines represent the least-squares fit of the data using a power law.}
    \label{fig:Iwate-Nikko}
    \end{figure}
%===============================================
%
    We first consider catalogs from two different volcanic regions (Iwate and Nikko) where the seismic activity is regarded as stationary. This means that the event rate is constant over the entire observation period. Clearly, the data presented in Fig.\ \ref{fig:Iwate-Nikko}(a) can be regarded as stationary. Moreover, no marked large earthquakes with $M > 5$ are reported in these two regions during the observation period. We further examine the validity of the GR law (Fig.\ \ref{fig:GR_IWATE_NIKKO}). This allows us to estimate the $b$-value as well as the completeness magnitude $M_c$ of the catalog (details in the \emph{supporting material}).

    Figure \ref{fig:Iwate-Nikko}(b) exhibits the probability density distribution $p(\tau)$ of IETs $\tau$ for these two catalogs on a double logarithmic scale. The logarithmically binned data for both the catalogs follows a decaying power law trend extending nearly three orders of magnitude at the intermediate range of inter-event times $\tau$. From the best fit of the data using a straight line on the log-log scale, we estimate the exponent value $\alpha=0.65(5)$ for Iwate and $0.61(3)$ for Nikko volcanoes.
    
%   %%%%%%%%% YAKEDAKE & HAKONE %%%%%%%%%
    \begin{figure}[t]
    \centering
    \includegraphics[width=0.98\linewidth]{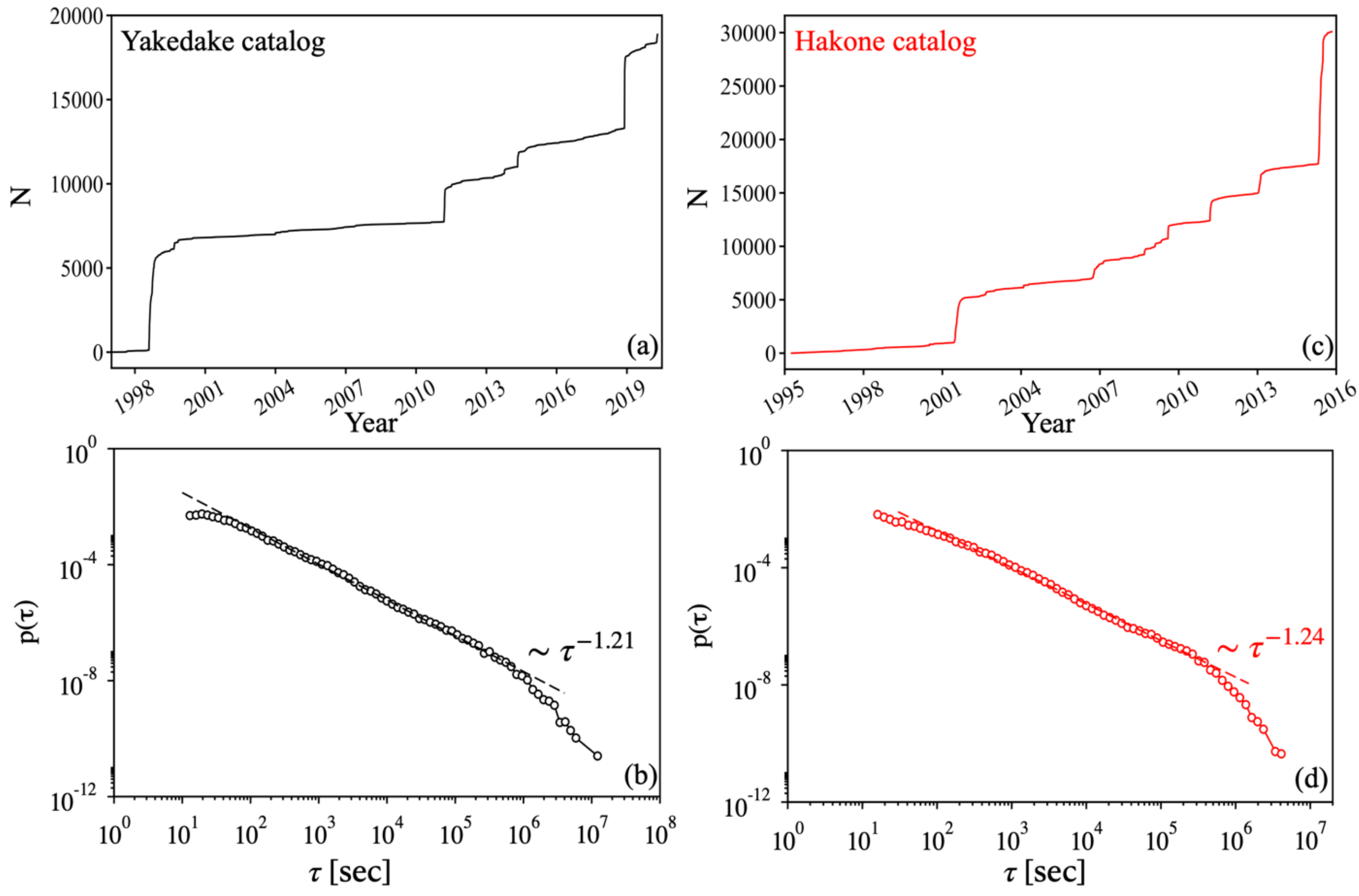}
    \caption{{\bf Non-stationary seismicity of active volcanoes with many bursts and distributions of IETs. (a, c)} The cumulative number of events as a function of time for the
    volcanic seismicity with $M \geqslant M_{\text{min}}$ recorded at Yakedake (left column) and Hakone (right column) within the observation window. {\bf (b, d)} Probability density distribution $p(\tau)$ of the IETs $\tau$ for the corresponding catalogs on a 
    log-log scale. The dashed line is the best fit to a power-law.}
    \label{fig:Yakedake-Hakone}
    \end{figure}
%    %%%%%%%%%%%%%%%%%%%%%%%%%%%%%%%%%%%%%%%
    
\textit{Non-stationary time series - Multiple bursts: Yakedake and Hakone - }
    To assess whether the IET distribution is affected by the nature of volcanic activity, we then examine relatively active volcanoes: Yakedake and Hakone. As shown in Figs.\ \ref{fig:Yakedake-Hakone}(a) and (c), the temporal activity of these two volcanoes is non-stationary with many bursts of earthquakes, which correspond to relatively minor hydrothermal unrest.
    %{\color{red} \bf{SK: Should we describe more details about the seismicity of these two volcanoes?}}
    
    Figs.\ \ref{fig:Yakedake-Hakone} (b) and (d) show the IET distribution $p(\tau)$ on a log-log scale with logarithmic bins. Power-law decay is again apparent in the intermediate time scale. Our estimate yields $\alpha=1.21(7)$ and $1.24(6)$ for Yakedake and Hakone volcanoes, respectively. These two values are again quite close. Importantly, they are significantly different from that for the stationary volcanic seismicity, $\approx 0.6$.
    By varying the cutoff magnitude $M_{\rm cut}$ from $M_{\text{min}}$ to $M_c$ (detailed descriptions in the \emph{supporting material}), we confirm that these exponents for Yakedake and Hakone ($\approx 1.2$) are insensitive to the missing events. This insensitivity is displayed in Fig.\ \ref{fig:distIET_Yakedake-Hakone_roubst}.

\textit{Non-stationary time series - Single Major Burst: Miyakejima - }
    We then consider the earthquake time series of a highly active volcano with a major eruption: i.e., Miyakejima volcano. The volcanic seismicity at Miyakejima is regarded as mostly stationary until the eruption in 2000. 

    \begin{figure}[t]
    \centering
    \begin{tabular}{c}
    \includegraphics[width=0.80\linewidth]{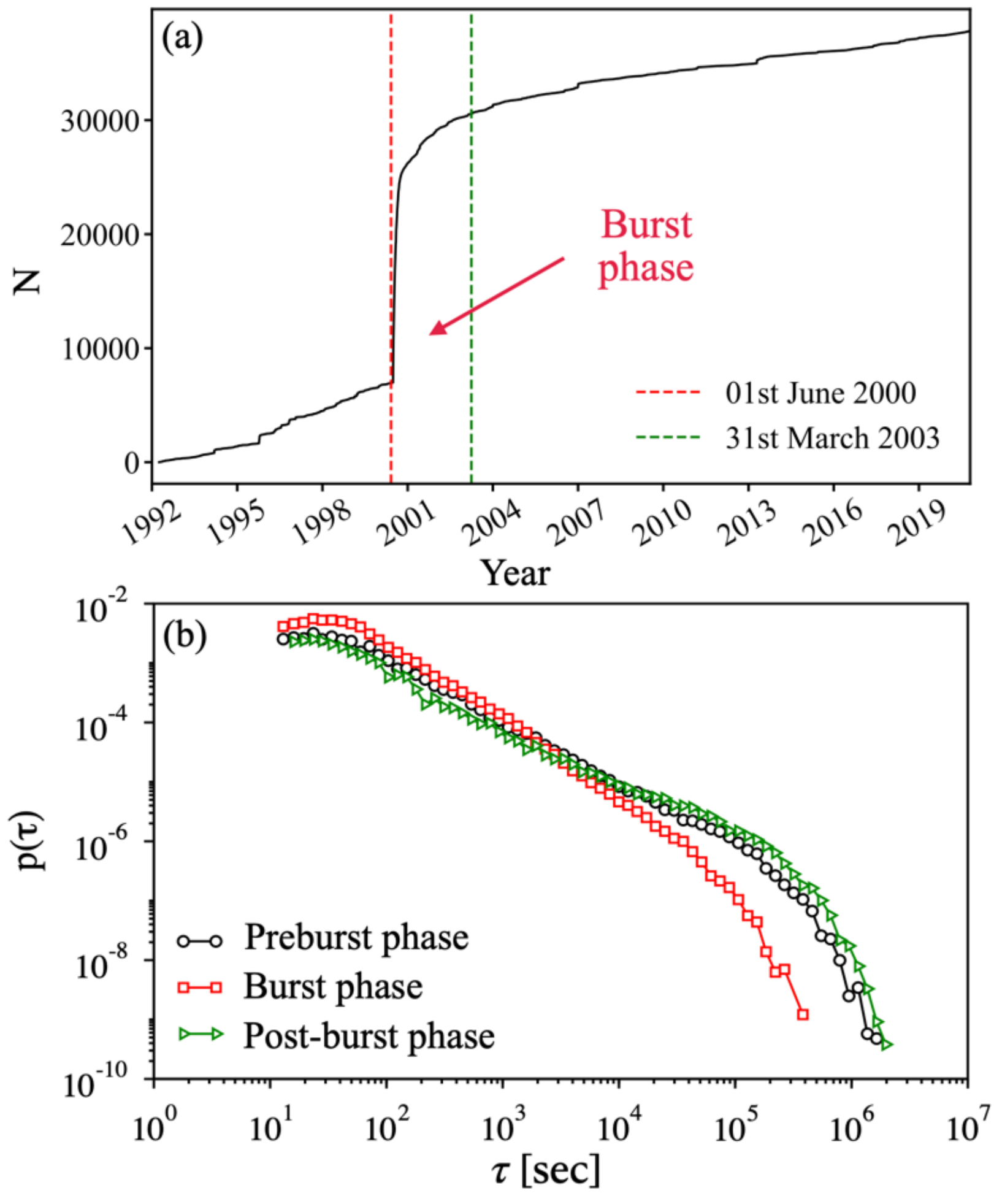}
    \end{tabular}
    \caption{{\bf Three distinct burst phases of the highly active Miyakejima volcano with non-stationary seismicity are well characterized by the exponent values of the IET distributions. (a)} The cumulative number of events as a function of time recorded between 1992-2020 with $M \geqslant 0$. {\bf (b)} The log-log plot of $p(\tau)$ vs.\ $\tau$ for the corresponding time series at different phases of the volcano: (black circles) preburst phase containing $N=6944$ events results the exponent $\alpha=1.03(6)$ for $10^2 < \tau < 2\times10^5$, (red squares) highly active bursty period with $N=23625$ events yields $\alpha=1.32(3)$ for $10^2 < \tau < 5\times10^4$, and (green triangles) post-burst steady state with $N=23625$ events yields $\alpha=0.84(3)$ for $10^2 < \tau < 5\times10^4$.}
    \label{fig:Miyakejima}
    \end{figure}
    
    On June 26, 2000, a swarm of volcanic earthquakes was first reported, followed by spatial migration of the earthquake hypocenters~\cite{JMA_Miyakejima2000,Ukawa2000}. From July 08, 2000, The summit subsidence with increased rate of seismicity, dike intrusions accompanied by the formation of caldera, and several phreatic eruptions were observed~\cite{Furuya2003,Nakada2005}. During the period July 01 - August 20, 2000, a total of $N=11674$ events with $M \geqslant 0$ are recorded, among which $622$ events have magnitude $M \geqslant 4$ and $50$ events with $M \geqslant 5$. The heavy volcanic activity continued until March 2003. The change in the seismicity and the signature of the strong non-stationary behavior during this period is clearly visible in the time series shown in Fig.\ \ref{fig:Miyakejima}(a).
    
    The time series of seismic activity in Miyakejima is thus classified into three different phases: preburst steady activity (April 1992 - May 2000), a burst of volcano-tectonic earthquakes (June 2000 - March 2003), and the post-burst steady activity (April 2003 - October 2020). See Fig.\ \ref{fig:TimeSeries_Miyakejima} for the typical time series in these three phases, exhibiting variability of volcanic events. We then analyze the IET distribution for each of the phases and examine the difference. The corresponding results are summarized in Fig.\ \ref{fig:Miyakejima}(b). In the intermediate time scale, we observe the power-law behavior over more than three orders of magnitude, but the exponent value is different for each stage of volcanic seismicity. We estimate the exponent $\alpha=1.03(6), 1.32(3)$, and $0.84(3)$ for the preburst, burst, and post-burst phases, respectively. 
    
    We again examine the robustness of these exponent values by varying the cutoff magnitude from $M_{\min}$ to $M_c$. The graphs are shown in Fig.\ \ref{fig:distIET_Miyakejima_roubst}. For the preburst and post-burst phases, we obtain the same exponent values within the standard error of the least-squares regression analysis. For the burst phase, we obtain the same value $\alpha \approx 1.33$, although the line does not appear to be quite straight in a certain range of $\tau$: $3 \times 10^3$ to $3\times 10^4$ sec.

    Although seismicity in the preburst phase appears to be stationary, the exponent value $(1.03)$ is significantly larger than the stationary cases of Iwate and Nikko $(\approx 0.65)$. This may suggest that the exponent for IET distribution could be a potential indicator of forthcoming eruption. Precursory physical processes before a major eruption, such as migration of highly pressurized magma, may have a notable influence on the exponent value. However, for the time being, we do not know the physical mechanism through which the exponent changes.
    
    The post-burst steady period exhibits slightly different exponent value $(\approx 0.85)$ from its stationary counterpart $(\approx 0.65)$. This is possibly because the physical state of volcano might not be the same after the eruption. Namely, a volcano typically undergoes the post-burst evolution processes accompanied by ground deformation and other related phenomena.

\begin{figure}[t]
    \centering
    \begin{tabular}{c}
    \includegraphics[width=0.80\linewidth]{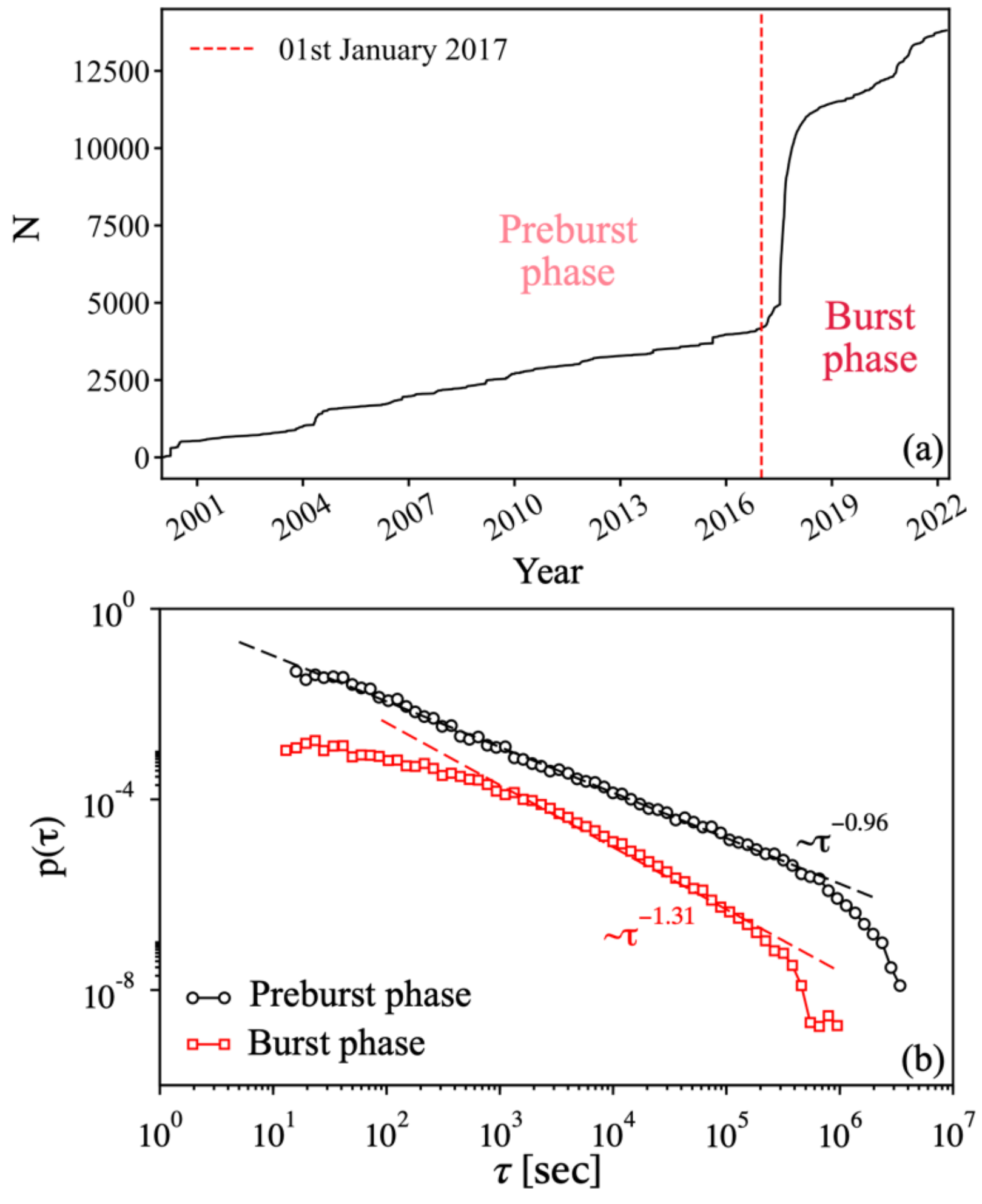}
    \end{tabular}
    \caption{{\bf The characteristic exponent values associated with the IET distributions recapitulate the volcanic seismicity of the preburst and burst phases for the Sakurajima volcano (a)} The cumulative number of events as a function of time with $M \geqslant 0.5$ recorded during the study period. {\bf (b)} The log-log plot of $p(\tau)$ vs.\ $\tau$: (black circles) For the preburst phase (2000-2016) with $N=4182$ events, and (red squares) for the highly active bursty phase (2017-2022) with $N=9627$ events. The dashed lines are the fit of the data using a power-law. For visual clarity, the data has been arbitrarily shifted in the vertical direction.}
    \label{fig:Sakurajima}
    \end{figure}

\textit{Non-stationary time series - Single Major Burst: Sakurajima - }
    To further validate the change of exponent value depending on the volcanic activity phase, we study Sakurajima volcano located in the nearly same tectonic settings. Similar to the case of Miyakejima, the volcanic seismicity at Sakurajima exhibited swarms induced by dike intrusions~\cite{Koike2021}, but consists of two phases: preburst and burst phases. See Fig.\ \ref{fig:Sakurajima}a. More details can be found in \cite{Koike2021,IGUCHI2019,Iguchi2022}. Again, the same pair of exponent values (within error bars) $\alpha=0.96(3)$ and $1.31(6)$ (Fig.\ \ref{fig:Sakurajima}b) describes the IET distributions for the preburst and burst phases, respectively. The Sakurajima case thus reinforces the hypothesis that the IET distribution exponent could be a potential indicator for volcanic eruptions.

\textit{Non-stationary time series - Single Major Burst: Kilauea - }
    We further analyze Kilauea volcano, where earthquake swarms and dike intrusion were observed during the 2018 eruption~\cite{Roman_Nature2021}. While the volcanoes considered so far are located in subduction zones, Kilauea is located on a completely different tectonic setting: i.e., hotspot. It is thus  interesting to compare the IET distribution for Kilauea with those in subduction zones. 
    
%    \tcr{[SK: We have events up to 60km in the catalog, but we decided to consider the events up to 30km. A major fraction of the events are shallow events. More precisely, 52682 no.\ of events are located within the depth of 10km among 53749 total events. See also Fig.\ \ref{fig:dist_depth}]} 
    
    \begin{figure}[t]
    \centering
    \begin{tabular}{c}
    \includegraphics[width=0.80\linewidth]{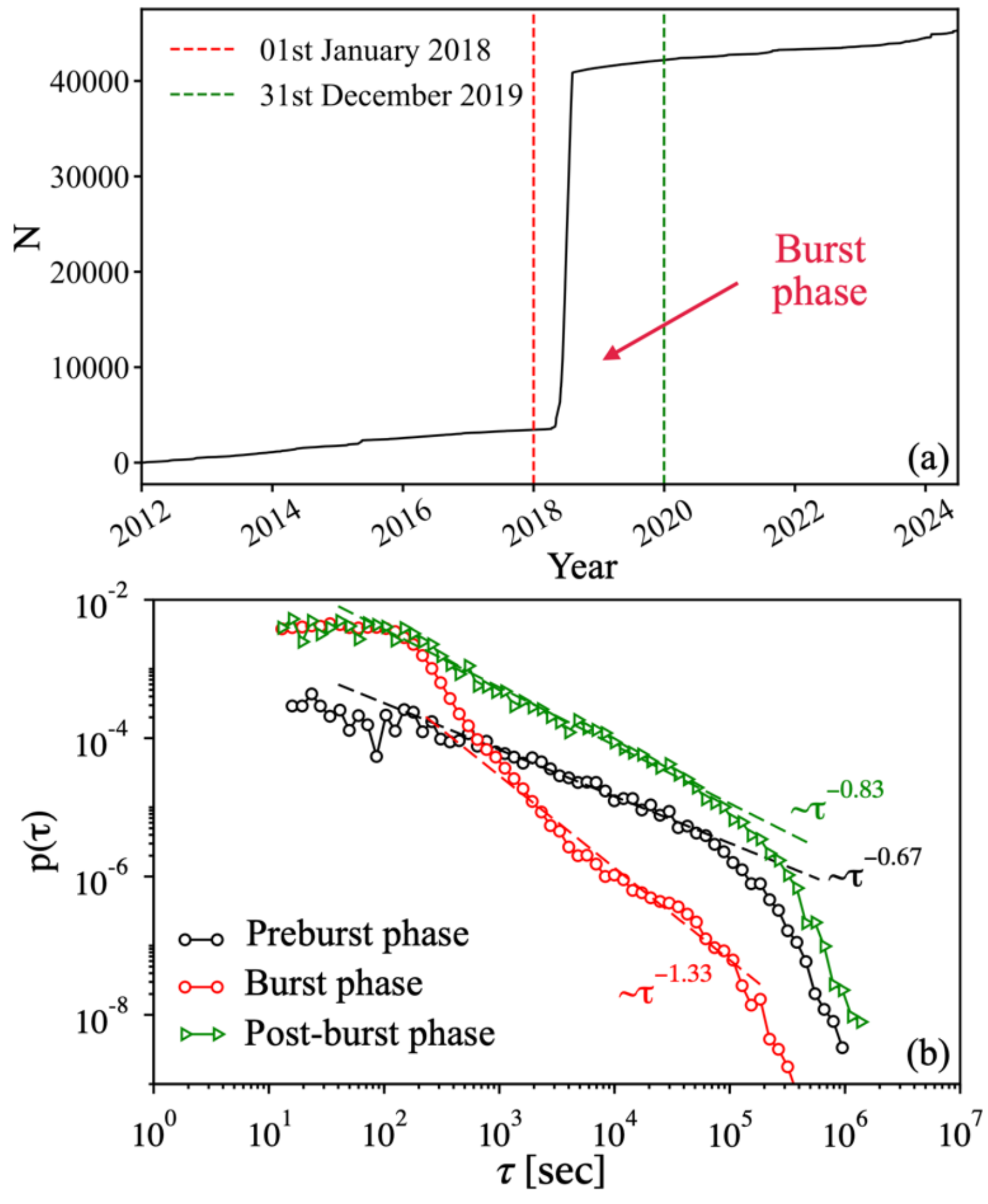}
    \end{tabular}
    \caption{{\bf The volcanic seismicity recorded at Kilauea exhibits the characteristic exponent values of the burst and post-burst phases. (a)} The cumulative number of events as a function of time with $M \geqslant M_c=1.8$ during the study period. {\bf (b)} The log-log plot of $p(\tau)$ vs.\ $\tau$: (black circles) For the preburst (2012-2017), (red squares) bursty burst (2018-2019), and (green triangles) post-burst (2020-2024) phases with $N=3440, 38750, $ and $3070$ events, respectively. For visual clarity, the data for post-burst phase has been arbitrarily shifted in the vertical direction.
    }
    \label{fig:Kilauea}
    \end{figure}
    
    The cumulative number of events is displayed as a function of time in Fig.\ \ref{fig:Kilauea}(a). Again, three phases are visible here: preburst, burst, and post-burst. However, the preburst time series appears to be more stationary than those in Miyakejima and Sakurajima. 
    
    We then compute the IET distribution for each phase. The IET distribution in the preburst phase is fitted with a power law and the exponent is estimated as $\alpha=0.67(3)$. This value is rather close to that for steady activity without eruption, such as the cases of Iwate and Nikko. Namely, volcanic earthquakes in the preburst phase of Kilauea is less precursory than the other cases.
    
    The IET distribution (Fig.\ \ref{fig:Kilauea}b) for burst phase has the exponent of $\alpha=1.33(6)$, common to other examples analyzed so far. The post-burst phase exhibits $\alpha=0.83(2)$. This value is also common to that for Miyakejima. 
    
    In summary, Kilauea exhibits common exponent values as those in Japan volcanoes except for the preburst phase. The difference in the preburst phase could be due to the nature of magma movement. Since the magma viscosity is low in Kilauea~\cite{Roman_Nature2021}, magma transport may be smoother and therefore rather similar to the stationary phase.   
    The common exponent values in burst and post-burst phases may be interesting, but at the same time puzzling considering many differences between Kilauea and the others analyzed here, such as the rheological and chemical properties of magmas, tectonic settings, and the structural details of conduit.

\textit{Conclusions - }We analyze the probability density distribution of inter-event time (IET) of volcanic earthquakes (mostly shallow, depth distribution in Fig.\ \ref{fig:dist_depth}) for several volcanoes. For the cases investigated here, all of them exhibit a power law decay $\tau^{-\alpha}$ in the intermediate time scale. The exponent $\alpha$ varies depending on the stage of volcanic seismicity. In particular, preburst seismicity may be characterized by a certain value of the exponent, $\alpha\simeq 1.0$. Since this is significantly different from the stationary phase ($0.6$ to $0.7$) and the burst phase ($\approx 1.3$), this could be interpreted as an index for imminent eruption with a burst of volcanic earthquakes.
 
 Interestingly, Kilauea does not exhibit this characteristic exponent for the preburst seismicity. Although we cannot clarify the reason for this difference for the time being, it may be attributed to the nature of magma transport rather than the tectonic settings (subduction zone or hotspot). This is because the volcanic seismicity should be most affected by the instability of magma transport in the crustal rocks. This can be elucidated if a volcano with high viscosity magmas (and hence erupts explosively) located in a hotspot. However, unfortunately, we do not have appropriate and sufficient data that allow the IET distribution analysis. This is an interesting future task to prove the universality of the present conclusion.

    \emph{Acknowledgements-}
    This study was supported by JSPS KAKENHI Grants JP21H05201 and 23K22416. Additional support from the MEXT Program: Data Creation and Utilization-Type Material Research and Development Project Grant JPMXP1122684766 and “Earthquake and Volcano Hazards Observation and Research Program” are also gratefully acknowledged.
    S. K. also acknowledges financial support from PNRR Grant CN\_00000013\_CN-HPC, M4C2I1.4, spoke 7, funded by NextGenerationEU.

%\bibliographystyle{apsrev4-2}
%\bibliography{IET}


%%%%%%%%%%%%%%%%%%%%%%%%%%%%%%%%%%%%%%%%%%%%%%%%%%%%%%%
% SUPPORTING INFORMATION
%%%%%%%%%%%%%%%%%%%%%%%%%%%%%%%%%%%%%%%%%%%%%%%%%%%%%%%
\newpage

%%%%
\widetext
\clearpage
\begin{center}
\textbf{\LARGE Supplementary information}\\
%\textbf{\Large  \\ \vspace*{1.5mm} Complex differences between volcanic earthquakes and non-volcanic tectonic tremors revealed by the inter-event time statistics} \\
\textbf{\Large  \\ \vspace*{1.5mm} Inter-event time statistics of earthquakes as a gauge of volcano activity} \\
\vspace*{5mm}
Sumanta Kundu, Anca Opris, Yosuke Aoki, and Takahiro Hatano
%Authors
\vspace*{10mm}
\end{center}
%\balancecolsandclearpage
%%%

%%%
\onecolumngrid
%%%

%

%%%%%%%%% Prefix a "S" to all equations, figures, tables and reset the counter %%%%%%%%%%
\setcounter{equation}{0}
\setcounter{figure}{0}
\setcounter{table}{0}
\setcounter{page}{1}
\setcounter{section}{0}
\setcounter{page}{1}
\makeatletter
\renewcommand{\theequation}{S\arabic{equation}}
\renewcommand{\thefigure}{S\arabic{figure}}
\renewcommand{\thetable}{S\arabic{table}}
\renewcommand{\thesection}{S\arabic{section}}
\renewcommand{\thepage}{S\arabic{page}}

\section{Methods}\label{sec11}
\subsection{Earthquake Catalogs}
%--------------------------------------------------------------------------------------------------------------------------
\begin{table*}[!htbp]
%\begin{table*}[t]
\caption{The summary of the earthquake catalogs analyzed for investigating the IET statistics. The data can be downloaded from the JMA website \cite{JMA} and the USGS website \cite{USGS}. The latitude and the longitude of the two diagonal vertices of the studied area are denoted as $(\theta_{\text{min}}, \phi_{\text{min}})$, and $(\theta_{\text{max}}, \phi_{\text{max}})$, respectively. The other columns are the period, the range of earthquake magnitudes $M \in [M_{\text{min}}, M_{\text{max}}]$, the maximum depth $d_{\text{max}}$ (in km), the total number of events $N$, the completeness magnitude $M_c$, and the number of events $N_c$ with magnitude $M \geqslant M_c$.
}
\begin{tabular*}{\linewidth}{l@{\extracolsep{\fill}}cccccccccr}
\hline \hline \vspace{-0.27cm} \\
Region  & $\theta_{\rm min}$ &  $\phi_{\rm min}$  & $\theta_{\rm max}$  & $\phi_{\rm max}$  & Period  & $[M_{\text{min}}, M_{\text{max}}]$ & $d_{\text{max}}$  & $N$   & $M_c$   & $N_c$ \\ \vspace{-0.27cm} \\ \hline \hline
Iwate     & 39.60 & 140.70    & 40.05 & 141.15    & 01/01/2000 -- 29/10/2020 & [0.0, 4.4]   & 28.0    & 5161  & 0.4 & 3836 \\
Nikko     & 36.50 & 139.25    & 36.85 & 139.65    & 01/04/2004 -- 31/12/2018 & [0.0, 4.8]   & 29.9   & 22815 & 0.4 & 15542 \\ \hline 
Yakedake  & 36.18 & 137.53    & 36.32 & 137.74    & 06/01/1997 -- 21/04/2020  & [0.0, 5.0]  & 29.5  & 17653 & 0.6   & 9740  \\
Hakone    & 35.15 & 138.90    & 35.35 & 139.10    & 06/04/1995 -- 31/10/2015  & [-1.9, 4.8]   & 8.0   & 30093 & 0.1   & 16279 \\ 
                Miyakejima    & 33.90 & 138.80    & 34.60  & 139.90 & 01/04/1992 -- 30/10/2020    & [0.0, 6.5]    & 30.0  & 37846 & 1.8   & 24960 \\
                Sakurajima & 31.103    & 130.40    & 31.695 & 130.89  & 01/01/2000 -- 30/03/2022  & [0.5, 4.4]    & 3.4   & 13809 & 0.5   & 13809 \\
                Kilauea   & 19.088    & -155.51   & 19.505    & -154.81   & 01/01/2012 -- 22/06/2024  & [0.1, 6.9]   & 30.0  & 53749  & 1.8 & 45260 \\ \hline \hline
\end{tabular*}
\label{tab:catalog}
\end{table*}
%--------------------------------------------------------------------------------------------------------------------------

   In this study, we analyze six volcanoes in Japan (Iwate, Nikko, Yakedake, Hakone, Miyakejima, and Sakurajima), and one volcano in Hawaii (Kilauea). For Hakone volcano, we use the earthquake catalog provided by the Hot Spring Research Institute \cite{Yukutake2015}. Otherwise we use the JMA unified earthquake catalog \cite{JMA} for volcanoes in Japan and the USGS catalog \cite{USGS} for Kilauea volcano. Table \ref{tab:catalog} summarizes the details of the catalogs.

\subsection{Definition of Inter-Event Time}
   In a given earthquake catalog, we first ensure that the data are ordered in the increasing 
   order in occurrence time. Then, the IETs are defined as
   \begin{equation}
       \tau_i = t_{i+1} - t_i,
   \end{equation}
   where $t_i$ represents the occurrence time of the $i$-th event in the catalog $(i = 1, 2, \cdots, N-1)$.
   Naturally, these time intervals exhibit large fluctuation depending on the seismicity and the size of selected region. In the main texts, the inter-event times are expressed in the unit of seconds.

\subsection{IET Distribution}
   The inter-event time distribution $p(\tau)$ is defined as the probability density function. Namely, $p(\tau)d\tau$ represents the empirical probability of having inter-event time between $\tau$ and $\tau+d\tau$: More precisely, the fraction of inter-event times having a value in $[\tau, \tau+d\tau]$. 

   Considering measurement uncertainties due to overlapping seismic waves from different events, we disregard inter-event times that are shorter than $10$ seconds.

   As presented below, the IET distribution develops the power-law regime in the intermediate timescale. Using the method of least-squares, the obtained IET distribution is fitted with a power law of $A\tau^{-\alpha}$. The two parameters $\alpha$ and $A$ are optimized by minimizing the standard error to obtain the best fit. The standard error in the exponent $\alpha$, representing the uncertainty in the slope of the best fitted straight line in the logarithmic space, is expressed as $\sqrt{\sigma_Y^2/\sum_i(\tau_i - \bar{\tau})^2}$, where $\sigma_Y^2$ is the residual variance, $\tau_i$ is the value of $\tau$ at each bin, and $\bar\tau$ is the average of $\tau_i$.
      
\subsection{Completeness Magnitude}
   The range of earthquake magnitude $M$ contained in the catalogs is denoted by [$M_{\min}, M_{\max}]$. Note that there are always missing records in earthquake catalogs, particularly events with smaller magnitudes. As a result, the GR law does not appear to hold below a certain magnitude. This magnitude is referred to as the completeness magnitude and denoted by $M_c$. In other words, a significant number of events are missed at $M < M_c$. Thus, any statistics below the completeness magnitude requires scrutiny.
   
\begin{figure}[!h]
    \centering
    \includegraphics[width=0.8\linewidth]{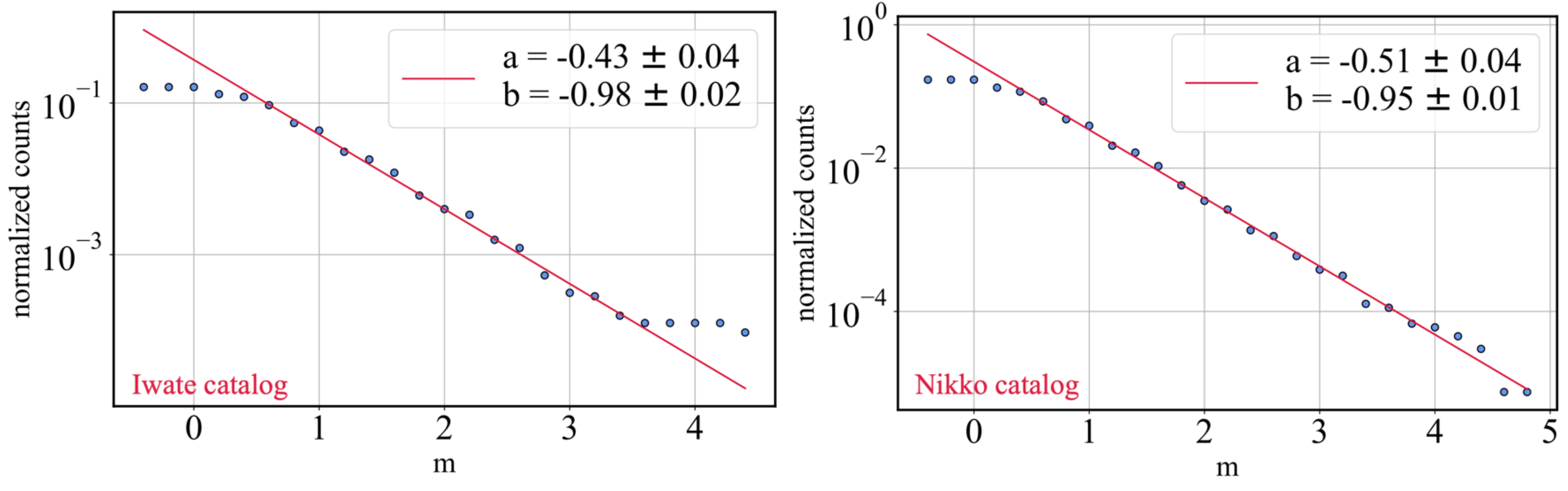}
    \caption{The frequency-magnitude distribution for volcanic earthquakes in Iwate and Nikko. The open circles represents the normalized cumulative number of events with magnitude $M \geqslant m$. The solid line represents the best fit of the data using the GR law for $M \geqslant M_c$.}
    \label{fig:GR_IWATE_NIKKO}
\end{figure}

\begin{figure}[!h]
    \centering
    \includegraphics[width=0.8\linewidth]{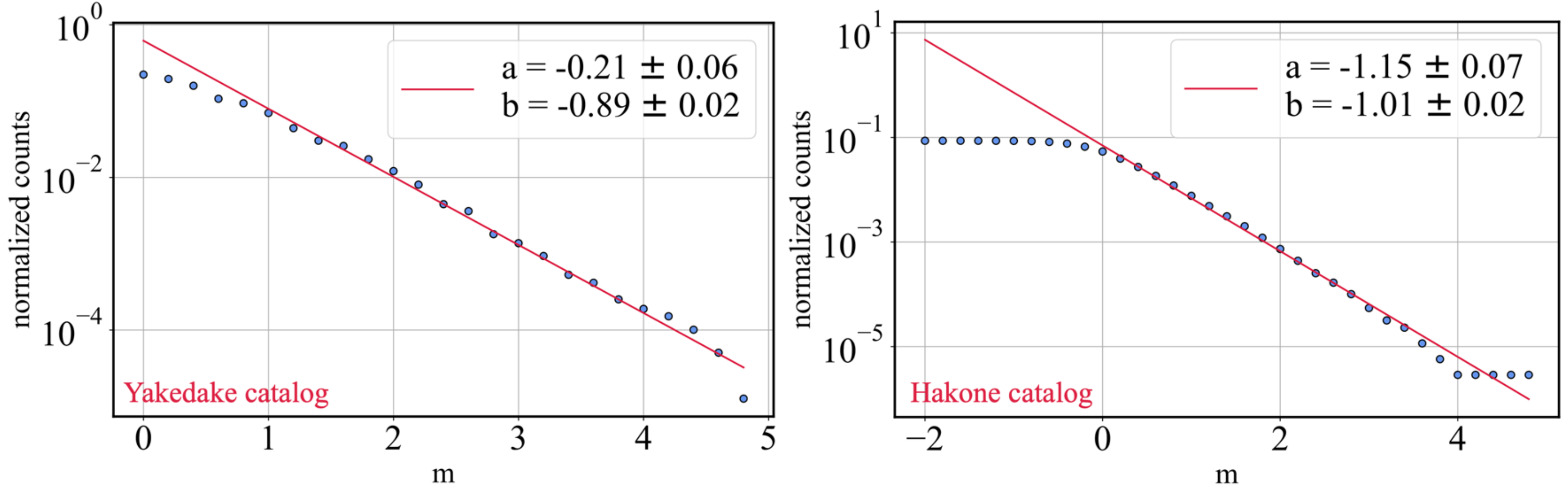}
    \caption{The frequency-magnitude distribution for volcanic earthquakes in Yakedake and Hakone. The open circles represents the normalized cumulative number of events with magnitude $M \geqslant m$. The solid line represents the best fit of the data using the GR law for $M \geqslant M_c$.}
    \label{fig:GR_Yakedake_Hakone}
\end{figure}

\begin{figure}[!h]
    \centering
    \includegraphics[width=0.98\linewidth]{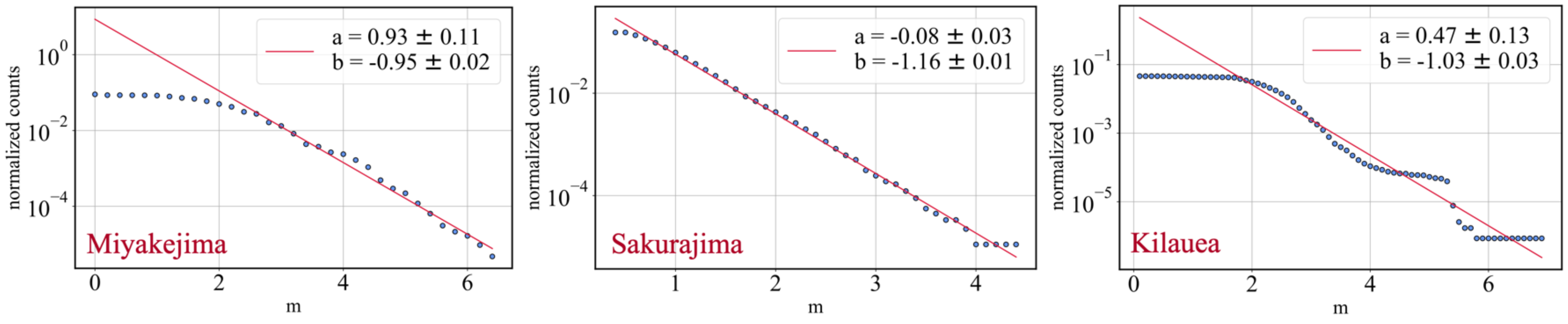}
    \caption{The frequency-magnitude distribution for volcanic earthquakes in Miyakejima, Sakurajima, and Kilauea volcanoes. The open circles represents the normalized cumulative number of events with magnitude $M \geqslant m$. The solid line represents the best fit of the data using the GR law for $M \geqslant M_c$.}
    \label{fig:GR_Miyakejima_Sakurajima_Kilauea}
\end{figure}

   The validity of the GR-law for the catalogs investigated in this study is summarized in Figs.\ \ref{fig:GR_IWATE_NIKKO}, \ref{fig:GR_Yakedake_Hakone}, and \ref{fig:GR_Miyakejima_Sakurajima_Kilauea}.
   %The concept of completeness magnitude is useful for most kinds of earthquakes except for tremors, for which the magnitude (and hence the GR law) is not well defined in a standard sense.
%

\subsection{Cutoff Magnitude}
   To clarify the influence of missing small events, in this study, we define the cutoff magnitude $M_{\rm cut}$, above which the analyses are made. Namely, we disregard any events below the cutoff magnitude. Note the difference between the cutoff magnitude $M_{\rm cut}$ and the completeness magnitude $M_c$. The completeness magnitude is unique in a given catalog, whereas the cutoff magnitude is an arbitrary parameter. We can thus check the robustness of the IET distributions (effect of missing events) by varying the cutoff magnitude $M_{\rm cut}$ from $M_{\min}$ to $M_c$. 

\subsection{Depth of Hypocenters}
\begin{figure}[!h]
    \centering
    \includegraphics[width=0.85\linewidth]{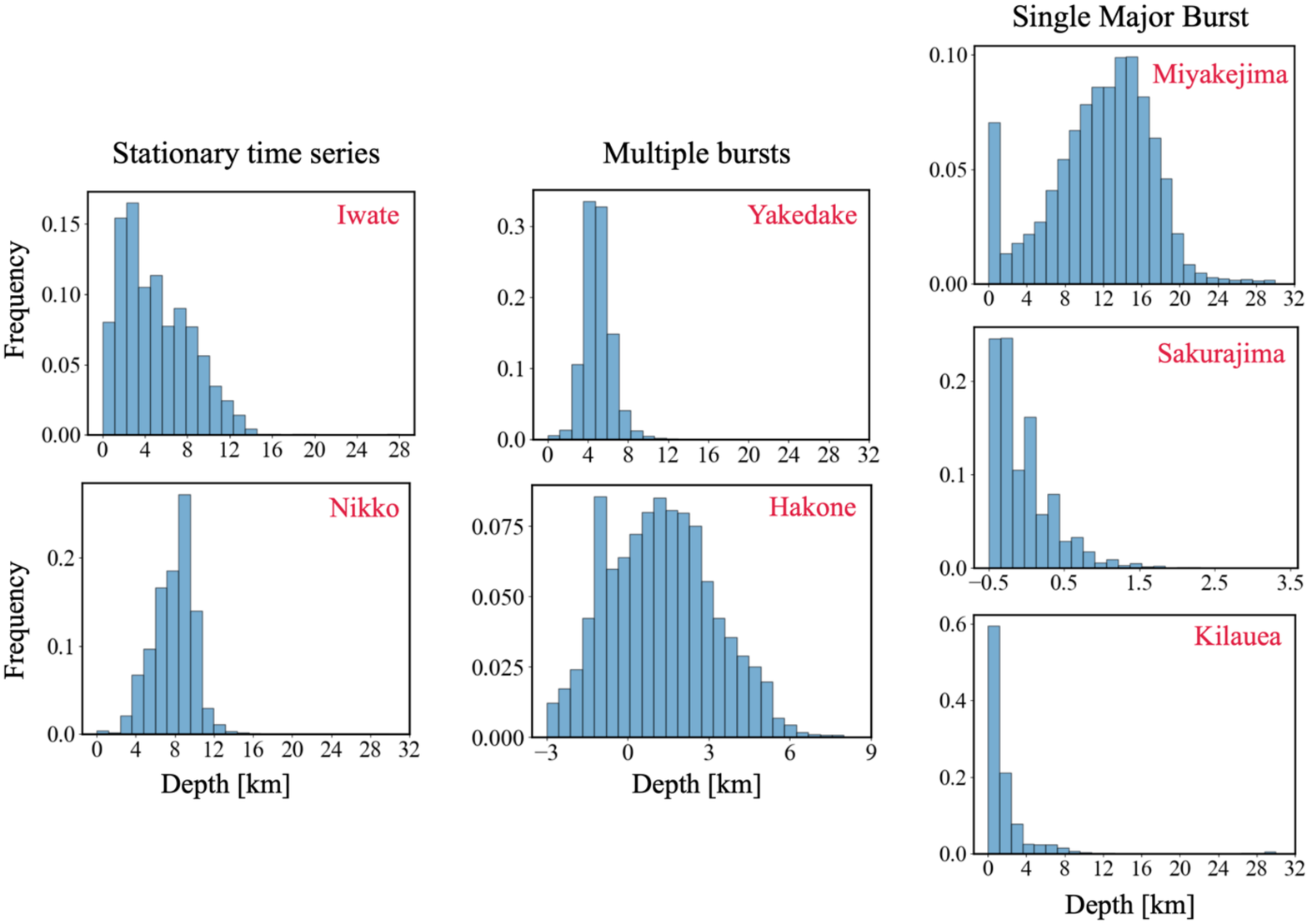}
    \caption{The depth distribution of the volcanic earthquakes with $M \geqslant M_{\rm min}$, and $d \leqslant d_{\rm max}$ for the analyzed catalogs. 
    }
    \label{fig:dist_depth}
\end{figure}

    Volcanic earthquakes occur mostly in the shallow crust beneath a volcano. We thus check the depth distribution of hypocenters for each catalog. As shown in Fig.~\ref{fig:dist_depth}, almost all earthquakes occur in the area shallower than $12$ km (except for Miyakejima). We can also confirm that the hypocenter distribution decreases monotonically and vanishes below a certain depth. Although there can be isolated deeper events, they may belong to a different category of earthquakes. We therefore believe that the catalogs covers the relevant volcanic earthquakes in a given area.

\begin{figure}[!h]
    \centering
    \includegraphics[width=0.8\linewidth]{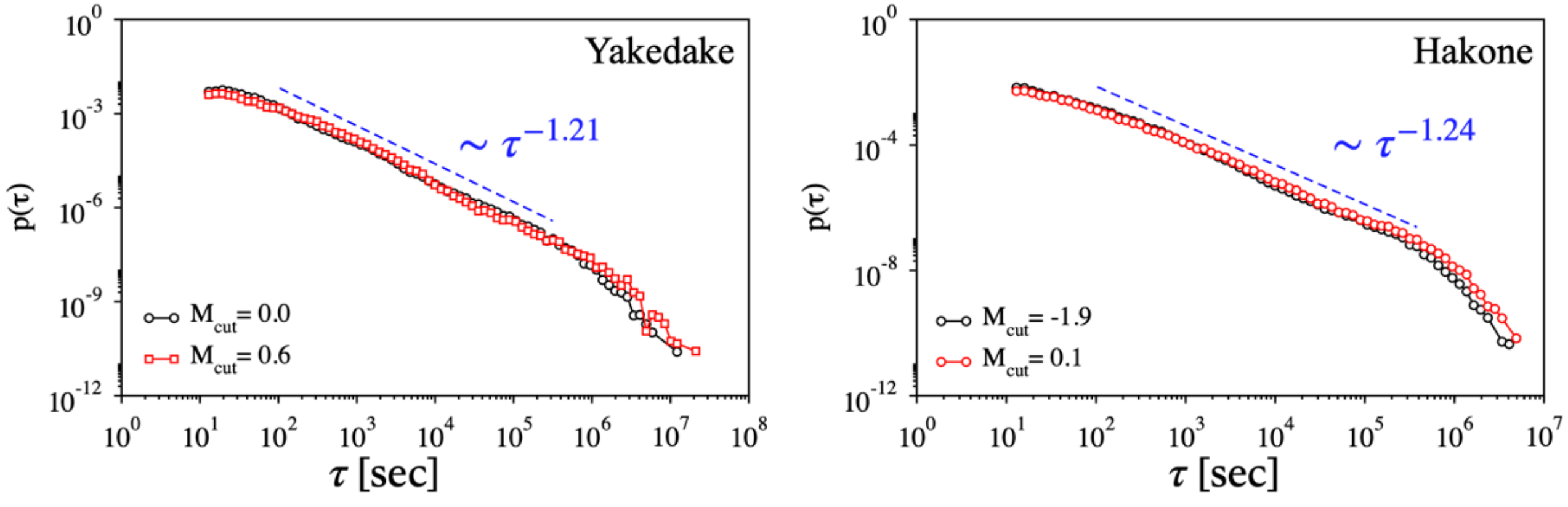}
    \caption{The log-log plot of $p(\tau)$ vs.\ $\tau$, demonstrating the robustness of the measured exponent values for Yakedake and Hakone volcanoes with $M_{\text{cut}}=M_{\text{min},~} \text{and~} M_c$ of the catalogs.
    }
    \label{fig:distIET_Yakedake-Hakone_roubst}
\end{figure}
\begin{figure}[!h]
    \centering
    \includegraphics[width=0.8\linewidth]{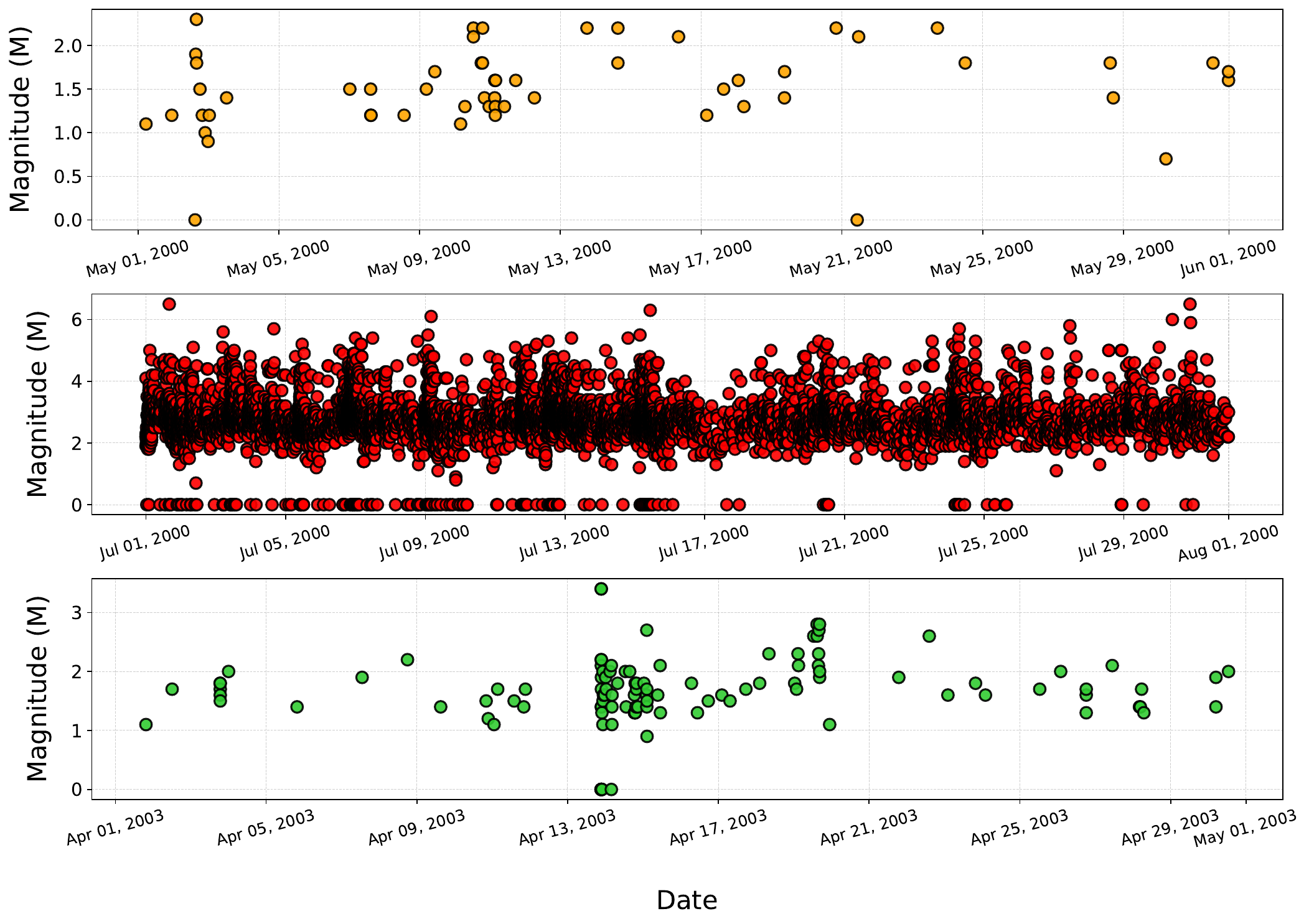}
    \caption{Time series of the Miyakejima volcano for the preburst (top), burst (middle), and post-burst (bottom) phases for the duration of one month with $M_{\text{cut}}=M_{\text{min}} (0.0)$, demonstrating the seismicity variability in three phases. The time series contains 57, 8738, and 112 events, respectively.
    }
    \label{fig:TimeSeries_Miyakejima}
\end{figure}
\begin{figure}[!h]
    \centering
    \includegraphics[width=0.98\linewidth]{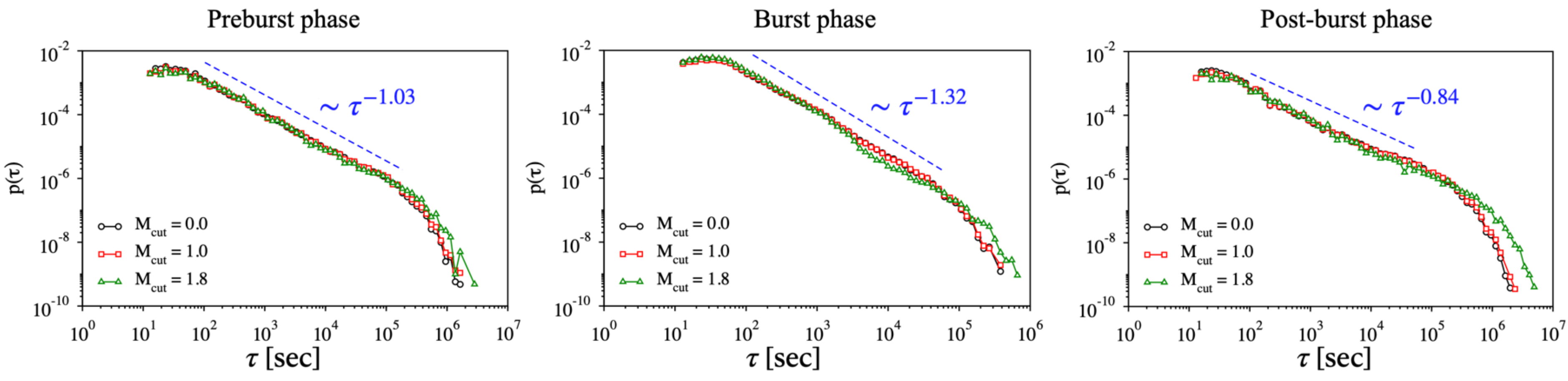}
    \caption{The log-log plot of $p(\tau)$ vs.\ $\tau$, demonstrating the robustness of the measured exponent values in the three different phases of the Miyakejima volcano. Three set of curves are for three different cutoff magnitudes $M_{\text{cut}}=M_{\text{min}} (0.0)$, $1.0, \text{and~} M_c (1.8)$.
    }
    \label{fig:distIET_Miyakejima_roubst}
\end{figure}

\end{document}